\RequirePackage{fix-cm}
\documentclass[smallextended]{svjour3}       % onecolumn (second format)
\smartqed  % flush right qed marks, e.g. at end of proof
\usepackage{graphicx}
\usepackage{diagbox}
\usepackage{multirow}
\usepackage{color}
\usepackage[square,sort,comma,numbers]{natbib}
\usepackage[colorlinks,linkcolor=red,anchorcolor=green,citecolor=blue
]{hyperref}
\usepackage{etoolbox}
\usepackage{threeparttable}
\usepackage{subfigure}
\usepackage[misc]{ifsym}

\journalname{Networks and Spatial Economics}

\newcommand{\tabincell}[2]{\begin{tabular}{@{}#1@{}}#2\end{tabular}}

\begin{document}

\title{Statistical Characteristics and Community analysis of Urban Road Networks}


\author{Wen-Long Shang \textsuperscript{1}, Huibo Bi \textsuperscript{*,1}, Yanyan Chen \textsuperscript{1}, Washington Ochieng \textsuperscript{2}}


\institute{   Wen-Long Shang \\
              shangwl\_imperial@bjut.edu.cn \\
           \\
           \Letter \quad Huibo Bi \\
              huibobi@bjut.edu.cn \\
              \\
              Yanyan Chen \\
              cdyan@bjut.edu.cn \\
              \\
              Washington Ochieng \\
              w.ochieng@imperial.ac.uk \\
              \\
              \textsuperscript{1} Beijing Key Laboratory of Traffic Engineering, College of Metropolitan Transportation, Beijing University of Technology. \\        %  \\
              \textsuperscript{2} Faculty of Engineering, Department of Civil and Environmental Engineering
}

\date{Received: date / Accepted: date}

\maketitle

\begin{abstract}
Urban road networks are typical complex systems, which are crucial to our society and economy. In this study, topological characteristics of a number of urban road networks based on purely physical roads rather than routes of vehicles or buses are investigated in order to discover underlying unique structural features, particularly compared to other types of transport networks. Based on these topological indices, correlations between topological indices and small-worldness of urban road networks are also explored. The finding shows that there is no significant small-worldness for urban road networks, which is apparently different from other transport networks. Following this, community detection of urban road networks is conducted. The results reveal that communities and hierarchy of urban road networks tend to follow a general nature rule.
\end{abstract}

\section{Introduction}
\label{intro}
Urban road networks (URNs) are vital in underpinning our society and economy (Neal \citeyear{neal2018urban}). Particularly, nowadays with the growing popularity of technologies related to smart city and cooperative vehicle infrastructure systems (CVIS), physical roads and its constituent URNs actually carrying most of urban traffics is increasingly receiving attention (Marshall and Gil et al. \citeyear{marshall2018street}). Meanwhile, URNs can be described as typical complex networked systems (Goh and Choi et al. \citeyear{goh2016complexity}), as there are large numbers of spatial components with complex structures and interactions between different components. Since Watts et al. (\citeyear{watts1998collective}) and Barabasi et al. (\citeyear{barabasi1999emergence}) proposed ``small world'' networks and scale-free networks respectively, complex network theory has been considered as a good tool to observe and analyse topological and structural features of networked systems (Ducruet and Beauguitte \citeyear{Ducruet2014complex}).

In fact, complex network theory has been applied to the study of various transport networks, such as public transport (B{\'o}ta and Gardner \citeyear{bota2017identifying}), air transport (Zhang and Derudder \citeyear{zhang2016dynamics}, Lozano and Guti{\'e}rrez \citeyear{Lozano2011airports}), railway transport and marine transport (Tsiotas and Polyzos \citeyear{tsiotas2015analyzing}), etc. Sienkiewicz and Holyst (\citeyear{sienkiewicz2005statistical}) explored the public transport network (PTN) systems of 22 cities in Poland. Topological characteristics of these cities, such as degree and degree distribution, average path length and cluster coefficients, tend to follow power-law distribution and exponential distribution, and these PTNs also match features of small-world networks. Following this, Xu and Hu et al. (\citeyear{xu2007scaling}) investigated the topological properties of three bus-transport networks in Beijing, Shanghai and Nanjing. The degree distribution in the three cities shows power laws in space $L$ (nodes refer to stations, and links denote that at least one route exists between two consecutive stations), while in space $P$ (nodes denote stations while links exist when two stations are shared by at least one route) that the cumulative degree distribution follows an exponential distribution. Similarly, Von Ferber and Holovatch et al. (\citeyear{von2009public}) explored the public transport networks (PTN) in 14 cities, and the results show that these PTNs have high clustering coefficients and relatively low average shortest paths, which are typical small-world characteristics. A directed and weighted bus transport network in Beijing also suggests that this bus network is a small-world network(Zheng and Chen et al. \citeyear{zheng2012analysis}). In order to examine the relationship between city size and network centrality, Derrible (\citeyear{derrible2012network}) investigated betweenness centrality of 28 metro systems across the world. Dimitrov and Ceder (\citeyear{dimitrov2016method}) examined the structure and topological characteristics of the bus network in Auckland, New Zealand based on the bus routes. Sun and Huang et al. (\citeyear{sun2018vulnerability}) analysed topological characteristics of Beijing rail transit network (BRTN), and vulnerable stations and cascading failures of BRTN were studied under different intentional attacks based on degree, betweenness and strength indices.

In addition to public transport networks, statistical characteristics of air networks(Guimera and Amaral \citeyear{guimera2004modeling}, Guimera and Mossa et al. \citeyear{guimera2005worldwide}, Guida and Maria \citeyear{guida2007topology}), railway networks(Li and Cai \citeyear{li2007empirical}, Guo and Cai \citeyear{guo2008degree}) and cargo ship networks(Hu and Zhu \citeyear{hu2009empirical}, Kaluza and Kolzsch et al. \citeyear{kaluza2010complex}) have also been explored based on complex network theory.
Although complex network theory has been frequently used to study many types of transport networks, complex network research related to urban road networks (URNs) consisting of road segments and junctions, which directly carry most of urban traffics, is very limited. Existing studies either investigated certain city road networks in dual representation (nodes refer to the set of edges sharing common attributes) so as to observe the growth of the city or model the congestion (Masucci and Stanilov et al. \citeyear{masucci2014exploring}, Zhan and Ukkusuri \citeyear{zhan2017dynamics}), or focused on the robustness and response of road networks under different attack strategies (Duan and Lu \citeyear{duan2014robustness}, Bellingeri and Bevacqua et al. \citeyear{bellingeri2018efficacy}). However, several questions related to URNs are still unclear: for instance, compared with other transport networks, what are the common statistical characteristics of URNs in different countries? How are the small worldness and underlying relationships between the statistical characteristics? How are the community and hierarchy features of URNs and what kind of rules they may follow, etc. To answer these questions, this paper investigates the topological characteristics of urban road networks from a number of cities in Europe and North America in order to explore substantial structure features of these networks. In addition, small-world analysis and community detection are also conducted.

The rest of this paper is organised as follows. Section \ref{background} introduces the widely-used topological indices in the field of network science. Section \ref{statanalysis} presents the statistical analysis of the urban road networks, and the small-world analysis is also conducted. Following this, correlation analysis between some topological indices is investigated in section \ref{correlationanalysis}. Section \ref{communitiesofURN} explores community detections of the urban road networks, and finally we draw conclusions in section \ref{conclusions}.

\section{Background and database}
\label{background}
In this paper, complex network theory is employed to explore topological characteristics of urban road networks (URNs), as it is a good tool to study very complex systems. Data associated with the URNs used in this study are obtained from a website, which is frequently used for transportation problems. Unlike typical Space $L$ and Space $P$ representations (Von Ferber and Holovatch et al. \citeyear{von2007network}), the nodes of an urban road network refer to the intersections of urban streets or roads, while the links represent the streets or road segments connecting nodes. Urban road networks can be divided into weighted or unweighted based on whether links are labelled with a weight, which can be the length of links or traffic flow on such links and so on. The weight of links depends on the purpose of research.

The URNs used in this paper mainly represent cities in Europe and North America, due to the fact that the limited scope of data sources. Specifically, these URNs are Austin, Chicago, Philadelphia, Anaheim, Winnipeg, Central Berlin, Barcelona, Terrassa and Hessen, and the size of the URNs varies, ranging from 416 nodes to 13389 nodes. Such information including the number of links is presented in Table \ref{tab:1}.

\begin{table}[htbp]
\caption{Urban road networks with different scales}
\label{tab:1}       % Give a unique label
\begin{tabular}{cccc}
\hline\noalign{\smallskip}
City & Nodes & Links & Locations \\
\noalign{\smallskip}\hline\noalign{\smallskip}
Austin & 7388 & 18961 & America \\
Chicago & 12982 & 39018 & America \\
Philadelphia & 13389 & 40003 & America \\
Anaheim & 416 & 914 & America \\
Winnipeg & 1052 & 2836 & America \\
Berlin & 12981 & 28376 & Europe \\
Barcelona & 1020 & 2522 & Europe \\
Terrassa & 1609 & 3264 & Europe \\
Hessen & 4660 & 6674 & Europe \\
\noalign{\smallskip}\hline
\end{tabular}
\end{table}

Table \ref{tab:2} summarises six topological indices used in this paper: Degree, Clustering Coefficient, Average Path Length, Closeness Centrality, Betweenness Centrality and Efficiency. We use these indices to analyse topological properties of URNs.

\begin{table}[htbp]
\caption{Six indices used for analysing urban road networks}
\label{tab:2}       % Give a unique label
\resizebox{\textwidth}{!}{
\begin{tabular}{ccm{4cm}} %m表示居中对齐，p表示顶部对其
\hline\noalign{\smallskip}
Indices & Equations & \multicolumn{1}{c}{Interpretation} \\ % \multicolumn{1}{c}{Interpretation} 单独把该单元格居中对齐
\noalign{\smallskip}\hline\noalign{\smallskip}
Degree & $k_i = \sum\limits_{j}\alpha_{ij}$ & where $\alpha_{ij}$ is the connection between node $i$ and node $j$; if $i$ and $j$ are connected $\alpha_{ij}$ is 1 otherwise it is 0. The degree of a network is the average of the degree of all its nodes. The degree distribution of the network $P(k)$ is a ratio of the number of nodes with degree $k$ to the number of all nodes and depicts the connective mechanism of networks. \\
\hline
Clustering Coefficient & $CC_i = \frac{EE_i}{k_i(k_i - 1)/2}$ &  where $EE_i$ is the actual number of edges between the neighbours of node $i$ (nodes connected with node $i$ are called its neighbours); $k_i$ is the number of the neighbours of node $i$ (Watts and Strogats \citeyear{watts1998collective}).  In social networks, this index tends to measure how close the friends of a given individual are. \\
\hline
Average Path Length & $APL = \frac{1}{n(n-1)/2} \sum\limits_{i>j}D_{ij}$ & where $D_{ij}$  is the length of the shortest path between $i$ and $j$. For un-weighted networks (i.e. every link has equal length of 1), $D_{ij}$ is the number of links of the shortest path between $i$ and $j$; while for weighted networks, assuming that the weight is simply physical distance, $D_{ij}$ is physical distance of shortest path from node $i$ to $j$. \\
\hline
Closeness centrality & $C_i = \frac{n-1}{\sum\limits_{v_j \in V, i \neq j}D_{ij}}$ & where $V$ is the set of nodes, $V=\{v_j:j=1,2,\cdots,n\}$, term $n$ is the number of nodes. \\
\hline
Betweenness centrality & $BC_i = \sum\limits_{j,Z \in V} \frac{NS_{jZ}(i)}{NS_{jZ}}$ & where $NS_{jZ}(i)$ is the number of shortest paths passing through the node $i$, and $NS_{jZ}$ is the total number of shortest paths between any pair of nodes. Weighted Betweenness centrality takes actual distance of the link as edge weight. \\
\hline
Efficiency & $E_{ff}(G) = \frac{1}{n(n-1)}\sum\limits_{i \neq j \in G} \frac{1}{D_{ij}}$  & This index was first proposed by (Crucitti and Latora et al. \citeyear{crucitti2003efficiency}) to explore the global efficiency of complex networks and looks very similar to $APL$; In this study, actual distance is assigned to the weight of a link. The greater $E_{ff}$, the better efficiency. \\
\noalign{\smallskip}\hline
\end{tabular}}
\end{table}

\section{Statistical analysis of the urban road networks}
\label{statanalysis}

Due to the fact that the complexity of urban road networks is partially arising from the topological characteristics and underlying structure features, thus statistical analysis plays a vital role in understanding such complexity and explore substantial mechanisms of urban road networks.

According to the background provided in Section \ref{background}, topological indices of these URNs are calculated in order to explore topological characteristics, and discover the underlying properties of the urban road networks (URNs) under consideration.

\subsection{Degree centrality}
\label{degreecentrality}

\begin{figure}[htbp]
\centering
\includegraphics[width=1\textwidth]{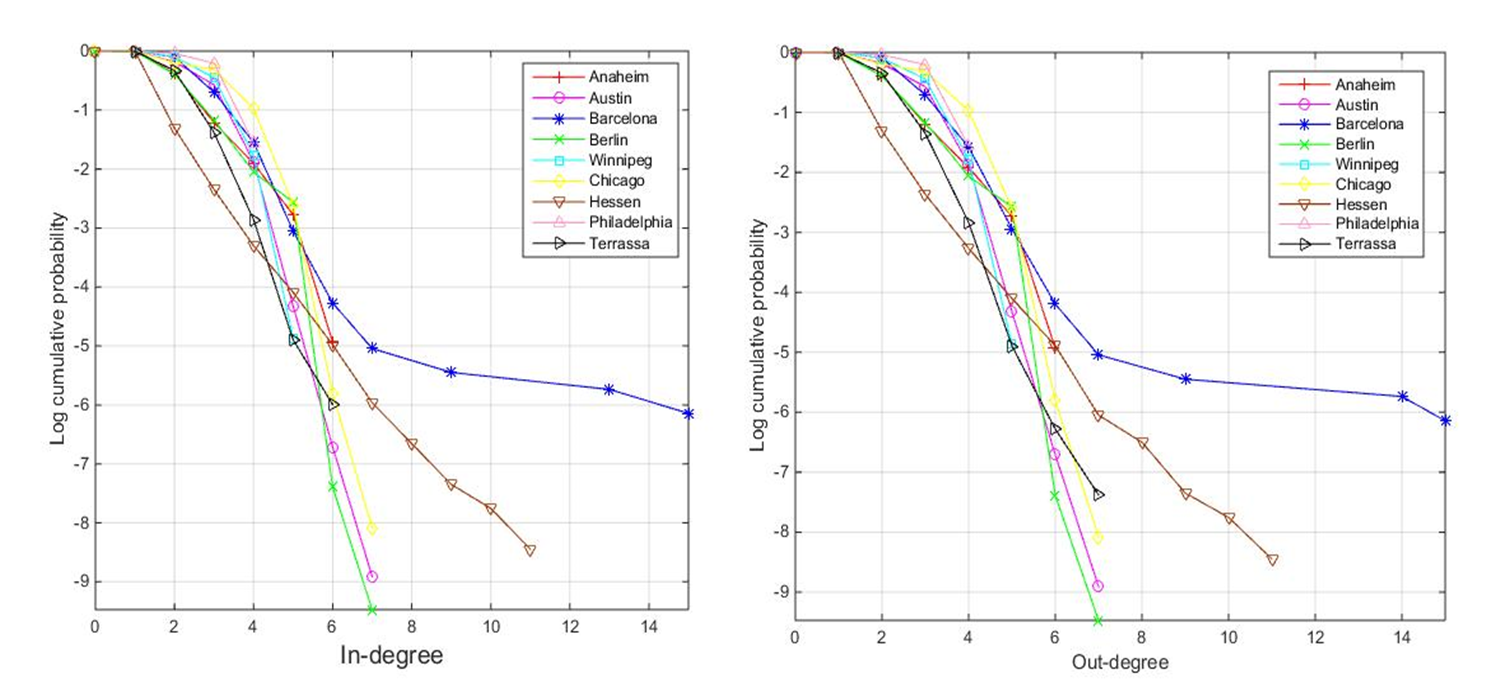}
\caption{Indegree and Outdegree of urban road networks.}
\label{fig:1}       % Give a unique label
\end{figure}

Degree is a seemingly simple yet very important index, since it reflects the node's connectivity and its importance in its vicinity. Therefore, this index is used to measure the local centrality of the networks. The degree centrality $k$ of a node $i$ in an undirected network (i.e. the links are undirecitonal) is the number of links connecting to this node (Barabasi and Albert \citeyear{barabasi1999emergence}). In realistic URNs, each intersection of streets/roads is regarded as a node, while the link represents the street/ road segments between the nodes. Here nine URNs used in this study are directional, therefore, the degree is divided into indegree (the number of links leading to a given node) and outdegree (the number of links leading away from a given node). The average indegree and outdegree of nine URNs have been summarised in Table \ref{tab:4}, and the degree distributions of nine URNs are illustrated in Fig. \ref{fig:1}.

In Fig. \ref{fig:1}, the y-axis represents logarithm of the cumulative probability $P(K \geq k)$, which is the probability that a randomly chosen URN node has indegree/outdegree equal to or larger than $k$. This section can observe that the range of indegree and outdegree for nine URNs are limited regardless of the size of the networks. For example, although the Philadelphia road network has 13389 nodes and 40003 links, the maximum indegree and outdegree are both 4. Among the nine URNs, Barcelona and Hessen have, respectively, the largest and second largest ranges of degree, despite their smaller sizes. The slope of the fitted curve in Fig. \ref{fig:1} depicts the speed of descent of the indegree and outdegree distribution curves, which reflects whether the majority of nodes of URNs are less connected.

Unlike the rest of the URNs, the indegree/outdegree of Barcelona has significant portions that are larger than 9, and the curves for Barcelona consist of two segments: the first one decreasing quickly while the second decreases slowly. Fig. \ref{fig:1} shows that most of nodes in these nine URNs are connected through a few nodes, and that the distributions of indegree and outdegree are qualitatively similar, which suggests that most of street/road segments in these nine URNs are bidirectional.

\begin{table}[htbp]
\caption{Summary of regression on six indices for nine URNs.}
\label{tab:3}       % Give a unique label
\begin{threeparttable}
\resizebox{\textwidth}{!}{
\begin{tabular}{ccccccccc}
\hline\noalign{\smallskip}
Indices & \multicolumn{2}{c}{\tabincell{c}{Indegree \\ distribution}} & \multicolumn{2}{c}{\tabincell{c}{Outdegree \\ Distribution}} & \multicolumn{2}{c}{\tabincell{c}{Weighted Closeness \\ Distribution}} & \multicolumn{2}{c}{\tabincell{c}{Weighted Betweenness \\ distribution}} \\
\noalign{\smallskip}\hline\noalign{\smallskip}
\multirow{2}{*}{\diagbox{City}{Parameters}{Function}} & \multicolumn{2}{c}{\tabincell{c}{\textbf{Power law} \\ $P(k_{in} \geq x) \propto x^{-\alpha}$}} & \multicolumn{2}{c}{\tabincell{c}{\textbf{Power law} \\ $P(k_{out} \geq x) \propto x^{-\beta}$}} & \multicolumn{2}{c}{\tabincell{c}{\textbf{Exponential} \\ $P(WC \geq x) \propto e^{\gamma x}$}} & \multicolumn{2}{c}{\tabincell{c}{\textbf{Power law} \\ $P(BC \geq x) \propto x^{- \lambda}$}} \\
 & $\alpha$ & $R^2$ & $\beta$ & $R^2$ & $\gamma$ & $R^2$ & $\lambda$ & $R^2$ \\
\hline
\textbf{Austin} & 2.549	& 0.99 & 2.55 & 0.99 & -135.1 & 0.88 & 0.6563 & 0.98 \\
\textbf{Chicago} & 3.253 & 0.99 & 3.254 & 0.99 & -262 & 0.92 & 0.6195 & 0.99 \\
\textbf{Philadelphia} & 6.981 & 0.99 & 7.002 & 0.99 & -118.9 & 0.88 & 0.5901 & 0.99 \\
\textbf{Anaheim} & 2.315 & 0.99 & 2.321 & 0.99 & -167300 & 0.99 & 0.7496 & 0.99 \\
\textbf{Winnipeg} & 4.559 & 0.99 & 4.539 & 0.99 & -48.93 & 0.90 & 0.7811 & 0.98 \\
\textbf{Berlin} & 2.879 & 0.97 & 2.87 & 0.97 & -72080 & 0.96 & 0.4645 & 0.99 \\
\multirow{2}{*}{\textbf{Barcelona*}} & 2.037 & 0.98 & 2.471 & 0.99 & -34.03 & 0.95 & 0.7355 & 0.99 \\
 & 0.2197 & 0.89 & 0.2193 & 0.91 & & & & \\
\textbf{Terrassa} & 2.002 & 0.98 & 1.776 & 0.98 & -20.27 & 0.99 & 0.7356 & 0.99 \\
\textbf{Hessen} & 0.9974 & 0.99 & 0.9998 & 0.99 & -112.7 & 0.99 & 0.7174 & 0.98 \\
\noalign{\smallskip}\hline
\end{tabular}}
\begin{tablenotes}
    \footnotesize
    \item The symbol ``*'' refers to that the indegree and outdegree distributions of Barcelona \\ follow a two-regime power law.
\end{tablenotes}
\end{threeparttable}
\end{table}

The second and the third columns in Table \ref{tab:3} show that most of the indegree and outdegree distributions fit the power law well. This result is consistent with previous work relating to public transport networks(Sienkiewicz and Holyst \citeyear{sienkiewicz2005statistical}, Von Ferber and Holovatch et al. \citeyear{von2009public}), worldwide maritime transportation networks represented in space $L$ (Hu and Zhu \citeyear{hu2009empirical}) and the Chinese bus-transport network(Xu and Hu et al. \citeyear{xu2007scaling}). It is worth noting that the indegree/outdegree distributions of some URNs also follow other distributions very well, for example, Anaheim, Berlin and Austin can also obey Gaussian distribution: $P(k \geq x) \propto e^{- (\frac{x - b_1}{c_1})^2}$, and the goodness of for indegree is 0.98, 0.95 and 0.96 respectively. Even so it is still enough to illustrate the decaying tendency of indegree/outdegree. In addition, the distributions of indegree and outdegree for these URNs show similarity, and this confirms that most of the links within the nine URNs are once again bidirectional. All node degree distributions are shown to follow the power law well: $P(k \geq x) \propto x^{-\mu}$ except that Barcelona follows a two-regime power law, and $\mu$ is the slope of the fitted curve (in the logarithmic coordinates) which depicts the speed of descent of the indegree and outdegree distribution curves.  A distribution curve with a steep slope suggests that the majority of nodes have smaller values of indegree/outdegree given that it is in the logarithmic coordinates; in other words, the URN overall is less well connected. Philadelphia, with the steepest slope among the nine URNs is overall less well connected than the other URNs.

\subsection{Clustering Coefficient}

\begin{figure}[htbp]
\centering
\includegraphics[width=0.8\textwidth]{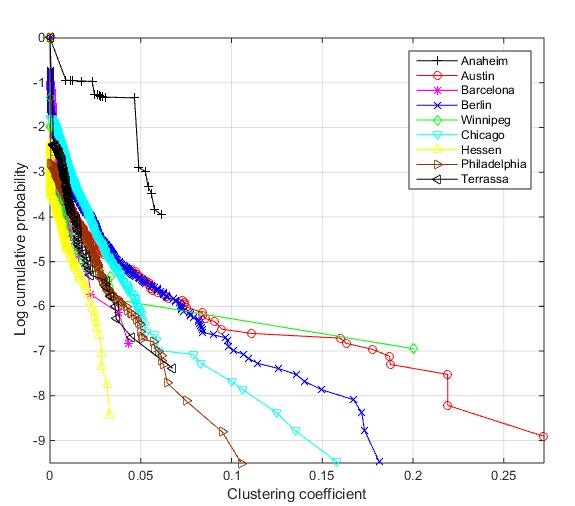}
\caption{Clustering coefficient of URNs.}
\label{fig:2}       % Give a unique label
\end{figure}

The clustering coefficient $(CC)$ is used to measure the local compactness of a network, and presents the clustering effect and local features of the network.  A larger $CC$ of a given node means the neighbours of the node are more likely to attract each other so that the local area of the network is more compact. According to the definition of $CC$ in Table \ref{tab:2}, the $CC$ for all URNs is calculated. Fig. \ref{fig:2} presents the log cumulative distribution $P(CC \geq x)$ of $CC$.

As can be seen from Fig. \ref{fig:2}, the distribution for Hessen has the steepest absolute slope, which means the road network of Hessen is the loosest compared to other eight cities due to the fact that most of its nodes have smaller values of $CC$, followed by Winnipeg and Philadelphia. The URN of Terrassa is looser than Barcelona, Chicago, Berlin and Austin, and the nodes of Anaheim tend to attract each other within the local range more than the other urban road networks in this study. The average $CC$ for the nine URNs has been summarised in Table \ref{tab:4}, which is normalized by the number of nodes, and also proves the viewpoint observed from Fig. \ref{fig:2}. The average $CC$ of these URNs is smaller compared to the Chinese airport network (Li and Cai \citeyear{li2004statistical}), the Italian airport network (Guida and Maria \citeyear{guida2007topology}), the world-wide air transportation network (Guimera and Mossa et al. \citeyear{guimera2005worldwide}) and the Boston subway network (Latora and Marchiorib \citeyear{latora2002boston}), possibly because the neighbouring nodes of a given node in the URNs are less likely to be connected to each other. This index is also able to be used to judge whether the network has small-world properties (Watts and Strogats \citeyear{watts1998collective}), and it will be depicted in detail later.

\subsection{Weighted Closeness Centrality (\textsl{WCC})}

\begin{figure}[htbp]
\centering
\includegraphics[width=0.8\textwidth]{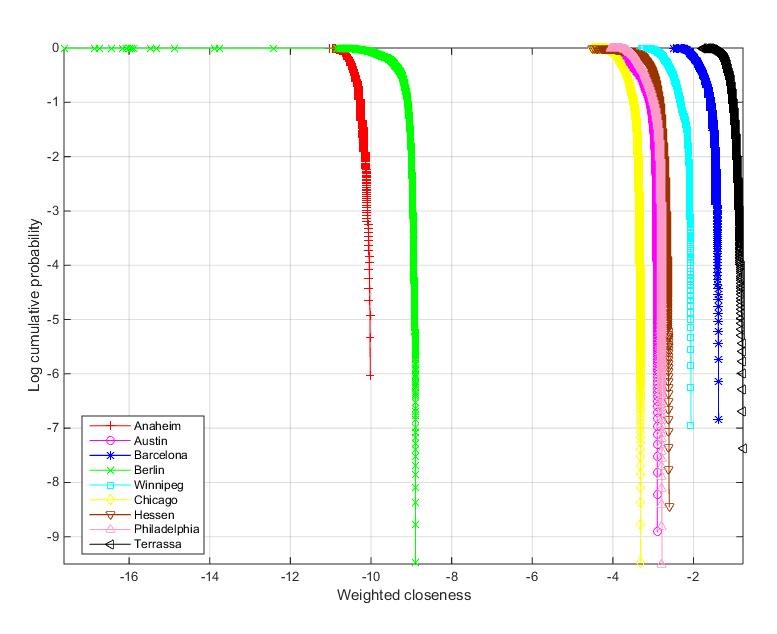}
\caption{Weighted closeness centrality of URNs.}
\label{fig:3}       % Give a unique label
\end{figure}

As previously introduced in Table \ref{tab:2}, closeness centrality (Sabidussi \citeyear{sabidussi1966centrality}) of a node $i$ is the inverse of the sum of the shortest path distance from node $i$ to other nodes, and it measures the accessibility of networks. In this study, the actual distances of links are taken as the link weights. According to function in Table \ref{tab:2}, the \textsl{WCC} for nine URNs are calculated. The cumulative distributions of \textsl{WCC} for these nine URNs are presented in Fig. \ref{fig:3} in a logarithmic scale. Given the differences in the size of these URNs, weighted closeness centrality is also normalized.

%\begin{figure}[htbp]
%\begin{minipage}[t]{0.5\linewidth}
%\centering
%\includegraphics[width=1\textwidth]{images/weightedclosenesscentrality.jpg}
%\caption{Weighted closeness centrality of URNs.}
%\label{fig:3}
%\end{minipage}%
%\begin{minipage}[t]{0.5\linewidth}
%\centering
%\includegraphics[width=1\textwidth]{images/weightedbetweennesscentrality.jpg}
%\caption{Weighted betweenness centrality of nine URNs.}
%\label{fig:4}
%\end{minipage}
%\end{figure}

As can be seen from Fig. \ref{fig:3}, all nodes of Anaheim and Berlin have relatively smaller \textsl{WCC}, which demonstrates that these two URNs are less accessible from a given node to other nodes, while Terrassa, Barcelona and Winnipeg have relatively large \textsl{WCC}. This conclusion is consistent with the result based on the average weighted closeness shown in Table \ref{tab:4}.

The \textsl{WCC} distribution is fitted for the nine URNs, as presented in Table \ref{tab:3}.

As observed from the fourth column in Table \ref{tab:3}, most URNs follow an exponential distribution, i.e. $P(WC \geq x) \propto e^{\gamma x}$, yet Austin, Philadelphia and Winnipeg do not fit this exponential distribution as well as other networks. From this distribution, it can see that Anaheim and Berlin have the largest $\gamma$ parameter values, which suggests that most of the nodes in these two URNs have smaller \textsl{WCC}, followed by Chicago, Austin, Philadelphia and Hessen, and Terrassa and Barcelona have a smaller $\alpha$ value. This is consistent with what Fig. \ref{fig:3} describes.

The average closeness centrality for nine URNs is summarised in Table \ref{tab:4}. The nodes with the largest \textsl{WCC} are generally thought as being located in the geographic centre of the city, since these nodes are easier to access than other nodes.

\begin{table}[htbp]
\caption{Indices of nine urban road networks}
\label{tab:4}       % Give a unique label
% resize the table to not extend the text width
\resizebox{\textwidth}{!}{
\begin{tabular}{ccccccccc}
\hline\noalign{\smallskip}
City & \tabincell{c}{Average \\ Indegree} & \tabincell{c}{Average \\ Outdegree} & \tabincell{c}{Average \\ Betweenness} & Closeness & \tabincell{c}{Clustering \\ coefficient} & $APL$ & Diameter & Efficiency \\
\noalign{\smallskip}\hline\noalign{\smallskip}
\textbf{Austin} & 2.5658 & 2.5658 & 0.0099 & 0.0393 & 0.0010 & 48.58 & 118 & 0.0284 \\
\textbf{Chicago} & 3.0062 & 3.0062 & 0.0053 & 0.0273 & 0.0017 & 44.71 & 111 & 0.0297 \\
\textbf{Philadelphia} & 2.9878 & 2.9878 & 0.0059 & 0.0446 & 0.0007 & 43.17 & 98 & 0.0290 \\
\textbf{Anaheim} & 2.1971 & 2.1971 & 0.0316 & 0.00003 & 0.0157 & 11.87 & 31 & 0.1098 \\
\textbf{Winnipeg} & 2.7269 & 2.7269 & 0.0253 & 0.0897 & 0.0007 & 18.85 & 40 & 0.069 \\
\textbf{Berlin} & 2.1855 & 2.1855 & 0.0075 & 0.0001 & 0.0017 & 50.09 & 116 & 0.0242 \\
\textbf{Barcelona} & 2.7118 & 2.7118 & 0.0203 & 0.1853 & 0.0009 & 14.00 & 31 & 0.0765 \\
\textbf{Terrassa} & 2.0362 & 2.0362 & 0.0181 & 0.3441 & 0.0009 & 25.09 & 64 & 0.0533 \\
\textbf{Hessen} & 1.4322 & 1.4322 & 0.0160 & 0.0516 & 0.0003 & 45.60 & 137 & 0.0282 \\
\noalign{\smallskip}\hline
\end{tabular}}
\end{table}

\subsection{Weighted Betweenness Centrality (\textsl{WBC})}

\begin{figure}[htbp]
\centering
\includegraphics[width=0.8\textwidth]{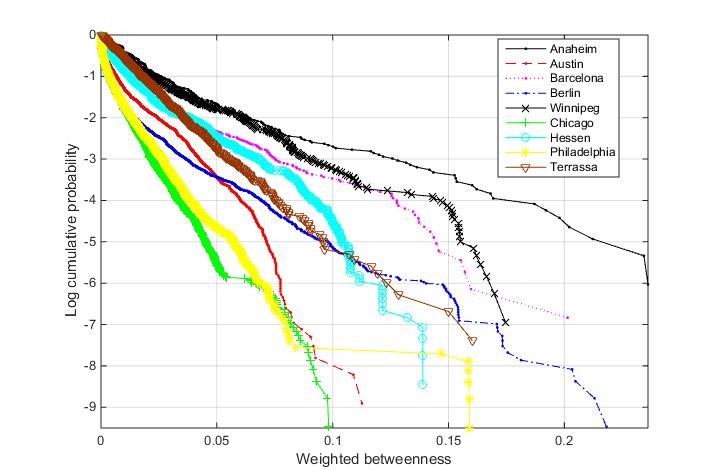}
\caption{Weighted betweenness centrality of nine URNs.}
\label{fig:4}       % Give a unique label
\end{figure}

Betweenness is a very important topological index in that it indicates that how frequently a given node is passed through by the shortest paths between all node pairs. In this study, the length of links is used as the edge weight, and \textsl{WBC} is calculated according to the function in Table \ref{tab:2}. The log cumulative distribution of \textsl{WBC} for the nine URNs is presented in Fig. \ref{fig:4}. To mitigate the effect of the size of different networks to betweenness, the weighted betweenness is normalized by dividing $n(n-1)$, where $n$ is the node number of the network. Fig. \ref{fig:4} exhibits that most of the nodes for the nine URNs have small \textsl{WBC}, and among all URNs, by intuition, Chicago overall shows the fastest decaying behaviour, followed by Philadelphia, and Austin; while Anaheim presents a relatively flat decaying tendency, followed by Winnipeg and Barcelona. As discussed, the larger betweenness centrality of a given node means the node is more important due to the fact that more of the shortest paths in the URN pass through this node. According to Sun, Wandelt et al. (\citeyear{sun2014topological}), a network is less robust if the proportion of nodes with large betweenness centrality is higher. The slopes of the curves in Fig. \ref{fig:4} show this proportion statistically, in other words, the steeper the slope is the more robust the URN is.

The log cumulative distribution of \textsl{WBC} can be also fitted by a power law function: $P(BC \geq x) \propto x^{- \lambda}$, and the results for this are summarised in Table \ref{tab:3}. In this case, normalised $BC \in (0,1)$, thus the larger $\lambda$ the slower the rate of decays.

Comparing all $\lambda$, Berlin exhibits the quickest decaying characteristic of \textsl{WBC} distribution, followed by Philadelphia, Chicago and Austin, which is against the intuitive observation. The reason could be that the distributions tend to overestimate or underestimate the decaying tendency of the curves in this case. The average weighted betweenness centrality (AWBC) of these URNs, however, which are presented in Table \ref{tab:4} show that Chicago has the smallest value, followed by Philadelphia, Berlin and Austin, and that Anaheim has the largest value, followed by Winnipeg and Barcelona. Statistically, the lower the $AWBC$ the network has the more robust the network is (Sun, Wandelt et al. \citeyear{sun2014topological}). The conclusion related to the robustness of these networks based on AWBC is therefore roughly consistent with that of observations from Fig. \ref{fig:4}.

WBC depicts how a given node of the network is passed by the shortest paths, and this information may be used for the allocation of important resources. Hence this index is sometimes used to assess the robustness and vulnerability of the network (Sun, Wandelt et al. \citeyear{sun2014topological}).

\subsection{Average path length $(APL)$, diameter and efficiency}

\begin{figure}[htbp]
\centering
\includegraphics[width=1\textwidth]{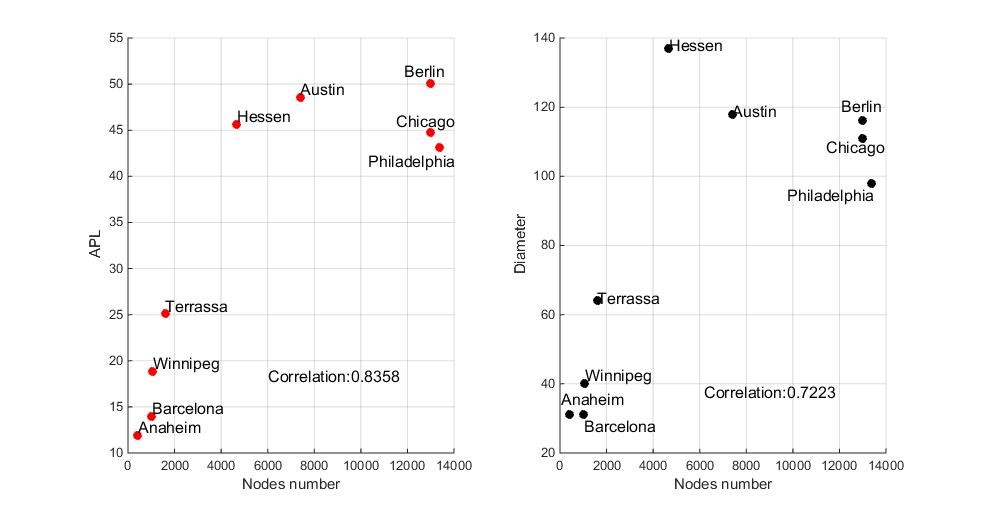}
\caption{Correlations between $APL$, Diameters and the size of nine URNs}
\label{fig:5}       % Give a unique label
\end{figure}

$APL$, also called the characteristic path length, can be employed to quantify the structural characteristics of networks (Watts and Strogats \citeyear{watts1998collective}). The diameter (\textsl{SD}) of the network is the maximum $D_{ij}$  for all node pairs. The $APL$ and the $SD$ are usually used to measure the efficiency and transport performance of the network. Here $APL$ can be categorized as weighted and un-weighted. The weighted $APL$ tends to measure the distance travelled from a given node to another node, while the un-weighted $APL$ assesses how many links need to be traversed from a given node to another. Both are able to assess the degree of accessibility of the URN. This thesis analyses un-weighted $APL$ to explore the structure of these URNs since there are no units for the distance data for the nine urban road networks, which means that comparisons of $APL$ between different URNs are meaningless. Furthermore, geodesic distance, namely, the number of links connecting a node pair is more focused in this study because of the homogenous characteristics of the physical links of each individual network. Accordingly, the $APL$ and the diameter for the nine URNs have been summarised in Table \ref{tab:4}. Anaheim has the smallest $APL$, which means the nodes of this network can be more efficiently connected by paths. This is followed Barcelona and Winnipeg, while the three networks with the largest $APL$ are Austin, Hessen and Berlin. The diameter and the $APL$ of the nine URNs are strongly correlated with the coefficient 0.9714 according to a Pearson correlation. In some cases, $APL$ and the diameter of networks are used to assess network performance (Albert, Jeong et al. \citeyear{albert2000error}, Crucitti, Latora et al. \citeyear{crucitti2003efficiency}), Crucitti and Latora et al. \citeyear{crucitti2004error}). The correlations between the size of networks, $APL$ and diameter are shown in Fig. \ref{fig:5}.

As can be seen from Fig. \ref{fig:5}, the Pearson correlations between $APL$, diameters and the size of networks are 0.8358 and 0.7223 respectively. This demonstrates that these two indices are positively proportional to the number of nodes to some extent.

In this study, $APL$ is calculated based on the function in Table \ref{tab:2}. If a network suffers from severe destruction, which may be generated by internal or external disruptions, so that the network is disconnected, an extremely large value is assigned to the disconnected links for the calculation of $APL$. While this approach has no effect on the ranking of  link importance, and comparisons the absolute value of $APL$ does not make sense. In order to better measure the performance of unconnected networks (Crucitti, Latora et al. \citeyear{crucitti2003efficiency}), an efficiency measure based on the inverse of the shortest path length is proposed in order to better observe the variations in the efficiency of different networks. According to the efficiency function presented in Table \ref{tab:2}, the results of efficiency for the nine URNs are calculated and also summarised in Table \ref{tab:4}. As can be seen, the results are consistent with the APL; for example, Anaheim has the smallest $APL$ (11.87) and, correspondently, its efficiency measure is the largest among all nine URNs, while Berlin has the largest $APL$ (50.09) and the smallest efficiency. In general, efficiency is able to be used to measure the performance of networks when suffering from disturbances.

\subsection{Small-worldness of urban road networks}

Small-world effect was first mentioned by Karkubthy (\citeyear{karinthy1929chains}) and mathematically presented by Pool and Kochen (\citeyear{de1978contacts}). The famous experiment related to small-worldness, ``six degrees of separation'', was conducted by Stanley Milgram in the 1960s (Milgram \citeyear{travers1967small}). Watts and Strogats (\citeyear{watts1998collective}) constructed networks which lie between regular and random networks, and named these ``small-world'' networks by analogy with Milgram's experiment. Small-world networks have $APL$ as small as $\textsl{APL}_{random}$ yet $CC \gg CC_{random}$ compared to random networks with the same number of nodes and edges, and a general feature of many complex networks is their small-world property (Watts and Strogats \citeyear{watts1998collective}), meaning that two nodes in the network can be connected with shorter paths. Following this, Newman (\citeyear{newman2003structure}) claimed that the value of the average shortest path for small world networks scales logarithmically or less with network size for a fixed mean degree. In order to explore whether these URNs have significant small-world properties, random networks which have the same number of nodes and the same average degree as these networks are generated using the \textbf{Pajek} software so as to calculate their average path length ($APL$) and clustering coefficient ($CC$). The comparisons between the generated random networks and the corresponding urban road networks are shown in Table \ref{tab:5}:

\begin{table}[htbp]
\caption{APL and CC comparisons between nine URNs and corresponding random networks of the same size.}
\label{tab:5}       % Give a unique label
\begin{tabular}{cccccc}
\hline\noalign{\smallskip}
City & $APL$ & $CC$ & $APL_{random}$ & $CC_{random}$ & $Ln(N)$ \\
\noalign{\smallskip}\hline\noalign{\smallskip}
\textbf{Austin} & 48.58 & 0.0010 & 9.2 & 0.0002 & 8.8940 \\
\textbf{Chicago} & 44.71 & 0.0017 & 8.6 & 0.0002 & 9.4713 \\
\textbf{Philadelphia} & 43.17 & 0.0007 & 8.6 & 0.0002 & 9.5022 \\
\textbf{Anaheim} & 11.87 & 0.0157 & 7.3 & 0.0032 & 6.0307 \\
\textbf{Winnipeg} &18.85 & 0.0007 & 6.9 & 0.0018 & 6.9584 \\
\textbf{Berlin} & 50.09 & 0.0017 & 12 & 0.0001 & 9.4712 \\
\textbf{Barcelona} & 14.00 & 0.0008 & 7.4 & 0.0013 & 6.9276 \\
\textbf{Terrassa} & 25.09 & 0.0009 & 9.3 & 0.0007 & 7.3834 \\
\textbf{Hessen} & 45.60 & 0.0003 & 18.9 & 0.0003 & 8.4468 \\
\noalign{\smallskip}\hline
\end{tabular}
\end{table}

As can be seen from Table \ref{tab:5}, the $APL$ of the nine real-life urban road networks (URNs) are much larger than those of the random networks with the same size. Although the $APL$ of Anaheim is the closest to that of the corresponding random network, it still takes 4.5 more. In the meantime, most of the $CCs$ for these nine URNs are larger than that for the corresponding random networks, but they are still in the same magnitude-level, except for Berlin. The $CCs$ of the URNs for Berlin and Winnipeg, meanwhile, are smaller than those for the corresponding random networks. In addition, the last column shows that $APL$ does not scale logarithmically with network size, $Ln(N)$, where $N$ is the number of nodes. The above observations show that the small-world property is not significant among these nine URNs.  In many studies of realistic networks, however, such as metabolic networks (Guimera and Sales-Pardo et al. \citeyear{guimera2007network}), the Chinese airport (Li and Cai \citeyear{li2004statistical}), the Italian airport network (Guida and Maria \citeyear{guida2007topology}), the world-wide air transportation network (Guimera and Mossa et al. \citeyear{guimera2005worldwide}), the Boston subway network (Latora and Marchiorib \citeyear{latora2002boston}), the worldwide maritime transportation network (Hu and Zhu \citeyear{hu2009empirical}) and public transport networks (Sienkiewicz and Ho{\l}yst \citeyear{sienkiewicz2005statistical}) small-world properties are evident.  This may be due to the fact that these URNs are spatial and that all links are directional, which may increase the average geodesic distance between any node pairs and reduce the clustering effect in any given local area. This finding shows that these URNs are different from other networked transport systems in the real world, and this may imply more profound impacts of other characteristics of these networks.

\section{Correlation analysis}
\label{correlationanalysis}

Many topological indices for the nine selected URNs have been calculated and analysed in Section \ref{statanalysis}. Based on this, a correlation between these topological indices can be conducted in order to explore the relationship between the topological indices and the underlying structural properties of these URNs. In addition, many interesting intuitive questions in the field of network science can be answered by correlation analyses of topological indices, for example, whether the most connected nodes are most central ones or those that are most compact at the local level (Guimera and Mossa et al. \citeyear{guimera2005worldwide}).

\subsection{Degree-betweenness correlation}

\begin{figure}[htbp]
\centering
\includegraphics[width=1\textwidth]{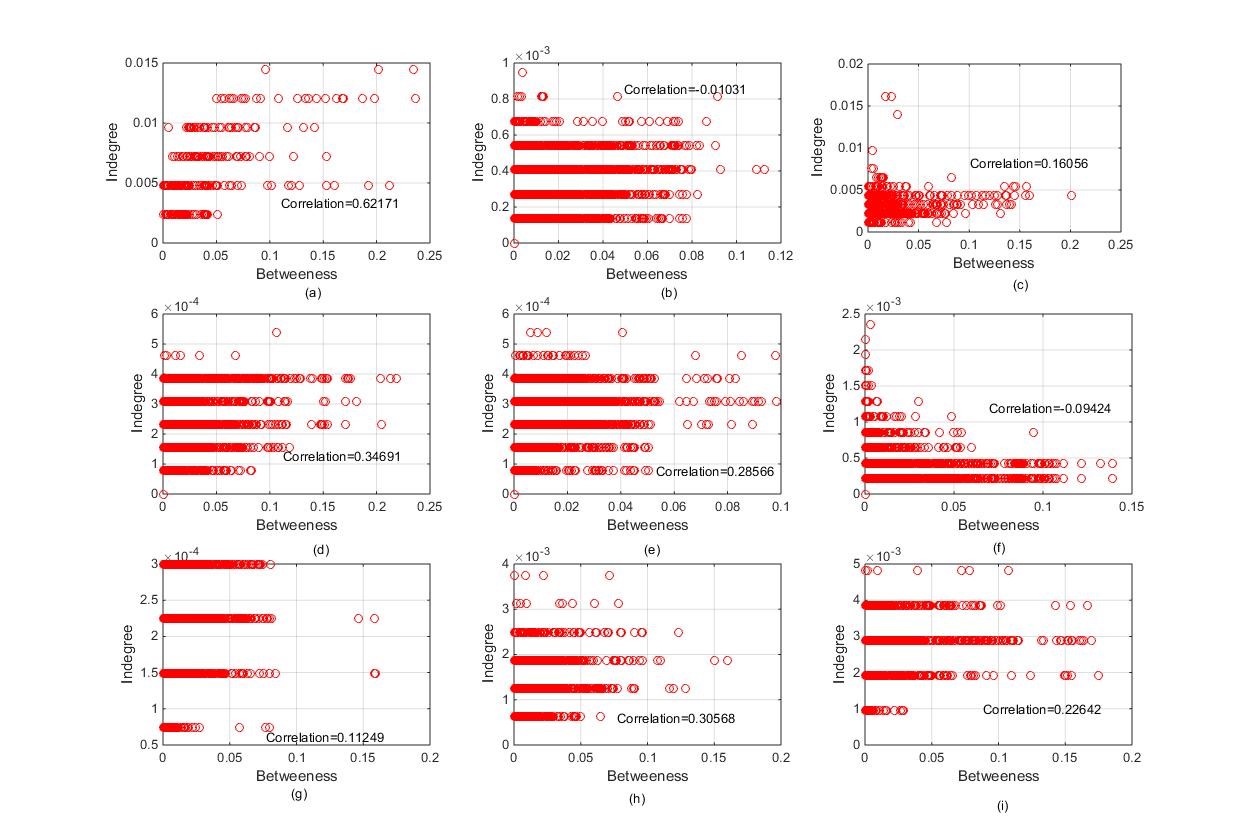}
\caption{The correlation between normalized indegree and betweenness: (a) Anaheim; (b) Austin; (c) Barcelona; (d) Berlin; (e) Chicago; (f) Hessen; (g) Philadelphia; (h) Terrassa; (i) Winnipeg.}
\label{fig:6}       % Give a unique label
\end{figure}

The nodes with large degree are most connected, and the nodes with large betweenness are most central in the network. Nodes with large/small betweenness and small/large degree are regarded as anomalies (Guimera and Mossa et al. \citeyear{guimera2005worldwide}). In fact, it is easy to design networks in which nodes have large betweenness and small degree. For example, if there are two communities connected by a single node with just two links, the node must have large betweenness due to the fact that it is frequently passed by routes which are going to one community from the other. This node plays an important role in such a network system. In order to check whether nodes with large degree have large betweenness for these URNs, this section calculates correlations between betweenness and indegree and plots them as shown in Fig. \ref{fig:6}.

\begin{figure}[htbp]
\centering
\includegraphics[width=1\textwidth]{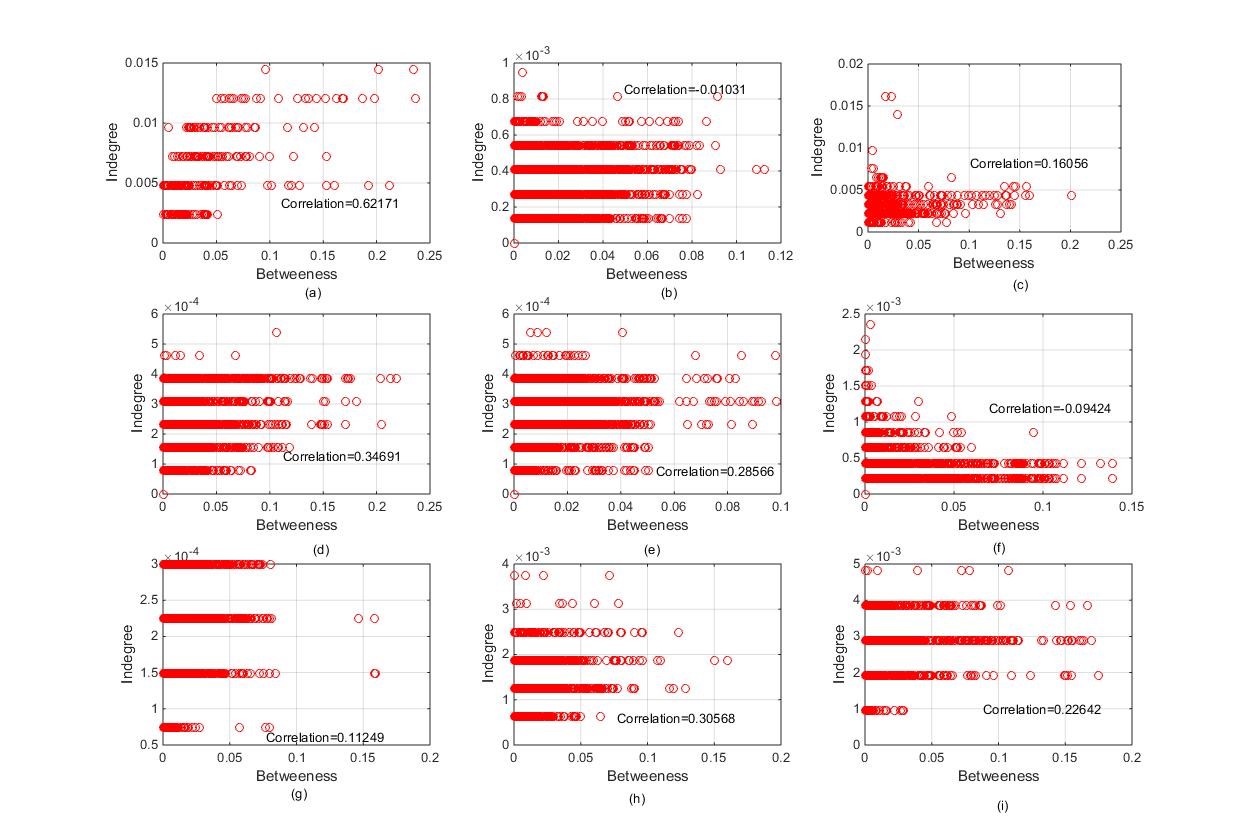}
\caption{The correlation between normalized  outdegree and betweenness: (a) Anaheim; (b) Austin; (c) Barcelona; (d) Berlin; (e) Chicago; (f) Hessen; (g) Philadelphia; (h) Terrassa; (i) Winnipeg}
\label{fig:7}       % Give a unique label
\end{figure}

Many complex networks, such as the Internet network (Vázquez and Pastor-Satorras et al. \citeyear{vazquez2002large}), show that nodes with large betweenness also have a large degree, while the worldwide airport network presents anomalous behaviour in this regard (Guimera and Mossa et al. \citeyear{guimera2005worldwide}). As can be seen from Fig. \ref{fig:6} the correlations between betweenness and indegree are weak, with neither presenting highly positive relationships or showing anomalous behaviour, but instead exhibiting more complex relationships. Anaheim has the highest correlation, 0.62, which means most of the city’s nodes are well connected as well as central.  Such relationships are almost non-existent for Austin, Barcelona, Hessen and Philadelphia, however, as shown by the low correlations (0.01, 0.16, 0.09 and 0.11, respectively). Such correlations may be caused by the distinct community characteristics of URNs. In addition, the correlations between outdegree and betweenness are identical to those between indegree and betweenness, as shown in Fig. \ref{fig:7}.

\subsection{Degree-clustering coefficient correlation}

\begin{figure}[htbp]
\centering
\includegraphics[width=1\textwidth]{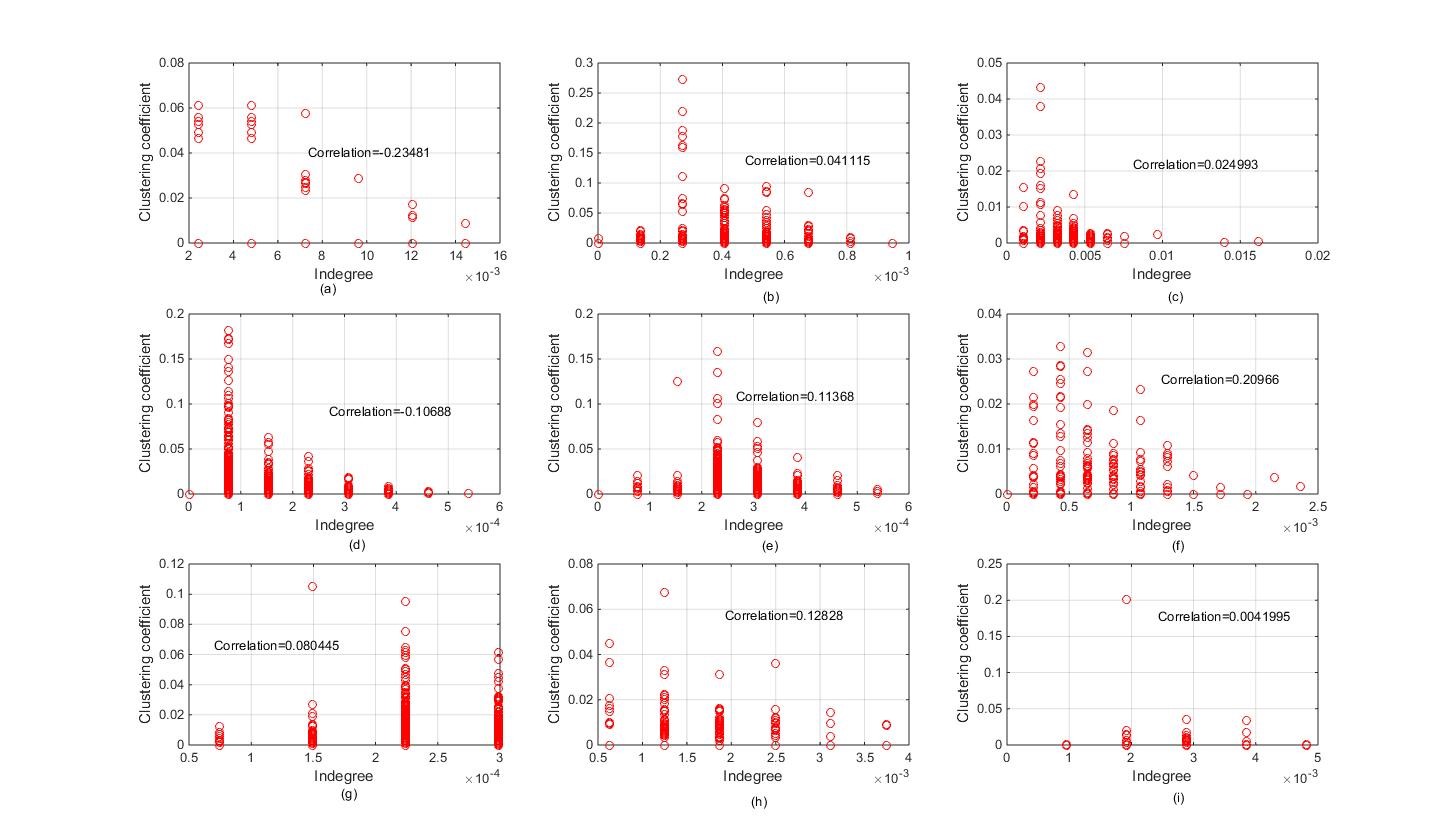}
\caption{The correlation between normalized indegree and clustering coefficient: (a) Anaheim; (b) Austin; (c) Barcelona; (d) Berlin; (e) Chicago; (f) Hessen; (g) Philadelphia; (h) Terrassa; (i) Winnipeg}
\label{fig:8}       % Give a unique label
\end{figure}

In order to examine whether the well-connected nodes in these URNs still have a large clustering coefficient $(CC)$, the correlations between $CC$ and indegree are calculated and plotted, as shown in Fig. \ref{fig:8}. It is not surprising that the correlations between them are very low, in fact all lower than 0.25, since this is identical to the analyses of $CC$. For these URNs, the neighbouring nodes are less connected with each other. The correlations below confirm this viewpoint.

The correlations between $CC$ and outdegree are similar to that between $CC$ and indegree, and against consistently very poor and lower than 0.25. The reason why urban road networks with many links have such poor correlations is that these networks are directional, and that the nodes in urban road networks tend to constitute quadrangles rather than triangles, which implies the neighbouring nodes of a given node are less likely to be connected with each other. This also explains why the clustering coefficients in these URNs are not very high.

\begin{figure}[htbp]
\centering
\includegraphics[width=1\textwidth]{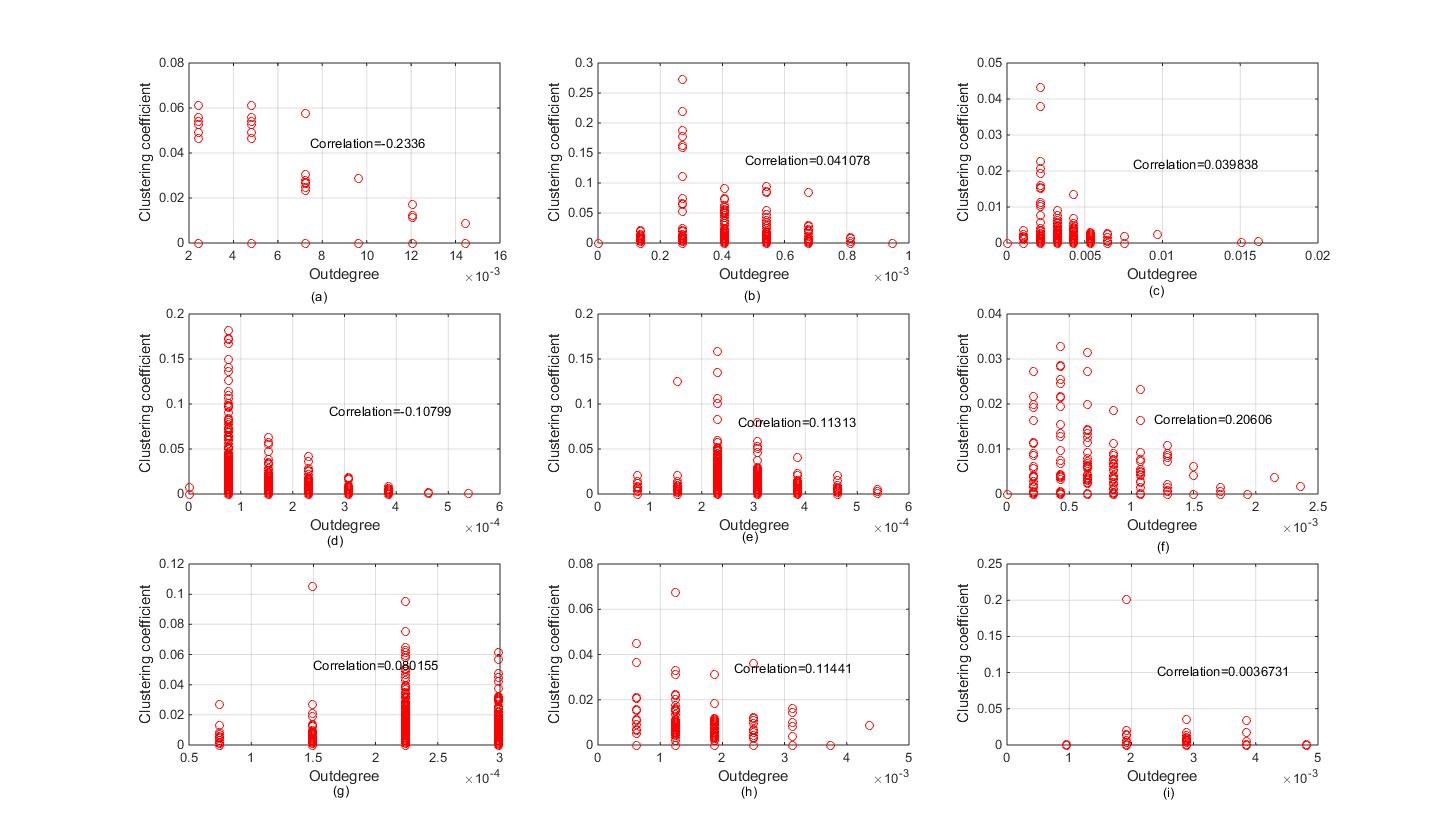}
\caption{The correlation between normalized outdegree and clustering coefficient: (a) Anaheim; (b) Austin; (c) Barcelona; (d) Berlin; (e) Chicago; (f) Hessen; (g) Philadelphia; (h) Terrassa; (i) Winnipeg}
\label{fig:9}       % Give a unique label
\end{figure}

\section{Communities of urban road networks}
\label{communitiesofURN}

The previous sections have performed topological analysis, small-worldness and correlation analysis between some topological indices. Building on these analyses, this section detects and analyses community structures within these networks in order to explore the topological characteristics of the URNs further.

The detection and analysis of the community structures within large-scale networks has been a focus of research in recent years. The quantitative definitions of communities within networks are diverse. Fortunato (\citeyear{fortunato2010community}) defined a community as a groups of nodes which play a similar role or share common features or properties within those groups. Lancichinetti and Radicchi et al. (\citeyear{lancichinetti2011finding}) regarded ``highly cohesive sub-graphs'' of networks as communities, clusters or modules. In biological and social networks (Fortunato \citeyear{fortunato2010community}, Illenberger et al. \citeyear{Illenberger2013}), community structure is a common feature, and it is thought of as a division of networks into densely connected subgroups (Newman and Peixoto \citeyear{newman2015generalized}). Lancichinetti and Radicchi et al. (\citeyear{lancichinetti2011finding}) proposed that a community is one pattern of node connections in realistic networks.

Detecting communities has many important applications in reality, such as identifying clusters of clients based on similar interests and geographical distance in order to provide better service (Krishnamurthy and Wang \citeyear{krishnamurthy2000network}), discovering communities of relationship networks between customers and online retailers so as to provide efficient purchase recommendations and marketing analyses (Reddy and Kitsuregawa et al. \citeyear{reddy2002graph}), and identifying clusters in large networks for data storage (Wu and Huberman \citeyear{wu2004finding}). In addition, the community structure of a network also offers a better way to understand networked systems more completely in that it is a powerful visualization tool to present the representation of networks rather than simply showing all the nodes and links of the network. In the transport field, in order to make clear the role of human collective behaviour phenomena across time and space in London transport, intertwined communities  of traffic across the whole city are investigated by Petri and Expert et al. (\citeyear{petri2013entangled}) to prove that this human spatial system is able to reach a self-organized critical state. Community detection has therefore been a hot topic in the modern science of complex systems (Fortunato \citeyear{fortunato2010community}).

Many methods and algorithms can be used to detect communities in graphs, such as graph partitioning, hierarchical clustering and spectral clustering (Fortunato \citeyear{fortunato2010community}), but they all have limitations in handling edge weights, edge directions or overlapping communities. In this study, Order Statistics Local Optimization Method (OSLOM), which is calculated based on the connectivity of networks and optimizes locally the statistical significance of clusters, is used to detect the community structure of these URNs. This was first proposed by Lancichinetti and Radicchi et al. (\citeyear{lancichinetti2011finding}) and is capable of handling directional and weighted networks, overlapping community structure, hierarchical structure, pseudo-communities, and so on, due to the fact that OSLOM focuses on optimizing locally the statistical significance of clusters. The detailed algorithm can be found in Lancichinetti and Radicchi et al. (\citeyear{lancichinetti2011finding})'s work.

\begin{figure}[htbp]
\centering
\includegraphics[width=1\textwidth]{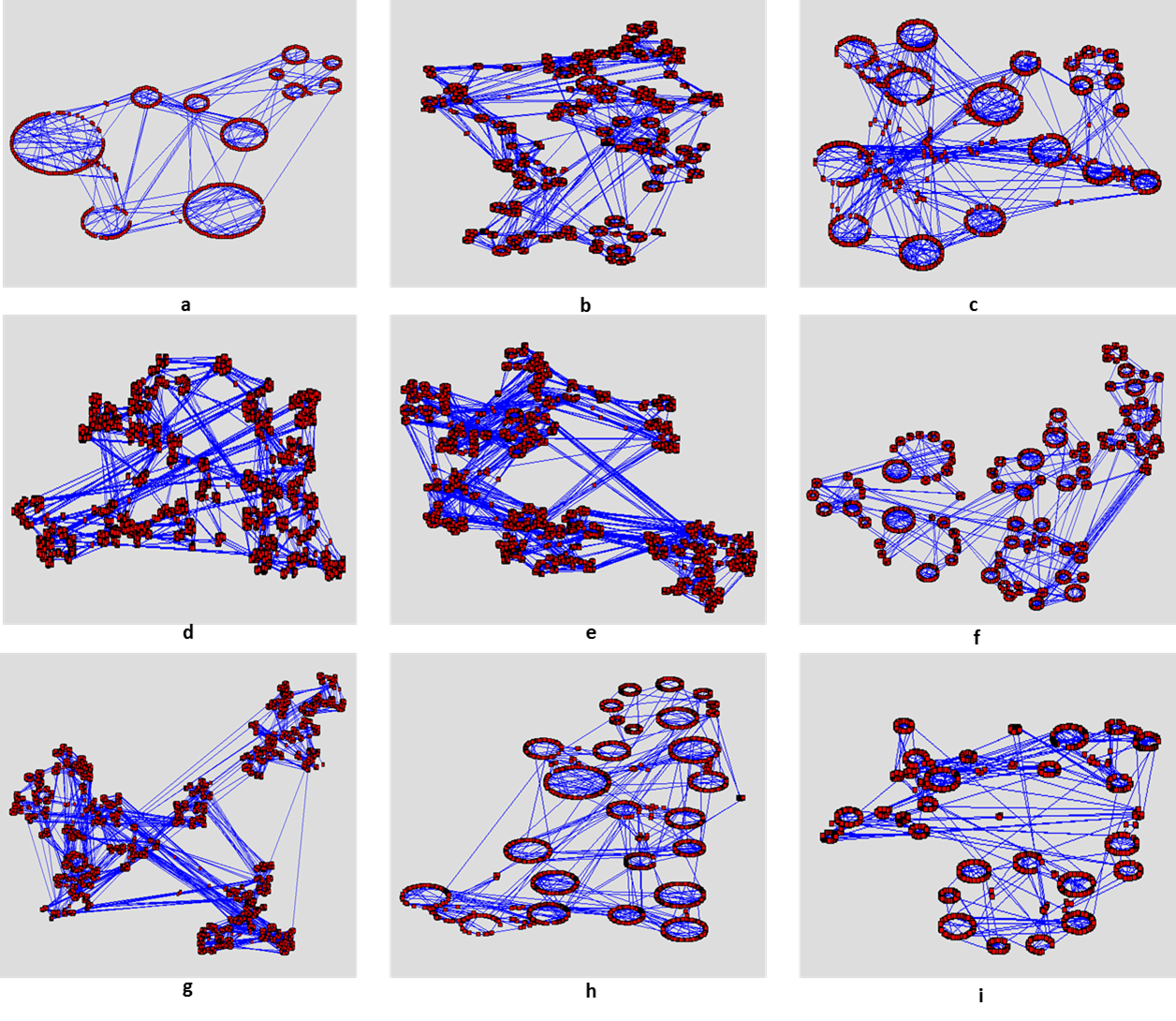}
\caption{Schematic graphs of communities for nine urban road networks: a. Anaheim; b. Austin; C. Barcelona; d. Berlin; e. Chicago; f. Hessen; g. Philadelphia; h. Terrassa; i. Winnipeg}
\label{fig:10}       % Give a unique label
\end{figure}

As a lifeline infrastructure, URNs can be regarded as one type of self-organized system, which means that they are similar to other systems in nature to some extent. In order to examine this natural property, community detection and analyses are conducted, based on the OSLOM method, and the results are shown in Fig. \ref{fig:10}.

The communities of all nine URNs are presented in Fig. \ref{fig:10}. The first impression of these communities is that the bigger the URNs the more the complex community structure they have. For example, Austin, Berlin, Chicago and Philadelphia are the larger networks, and the number of communities for these URNs are apparently more than those in the smaller networks. In addition, it seems that these bigger URNs have a significant hierarchical structure, namely, the bigger clusters consist of several smaller ones. The smaller networks, such as Anaheim, Terrassa and Winnipeg, meanwhile, present a clearer community structure. In addition, some nodes which look like not belonging to any communities in these URNs are overlapping ones, which suggests that these nodes are shared by several communities, and the nodes may assist the mutual communications for these communities, thus these ``sharing nodes'' may have great impact on the robustness and vulnerability of the whole network.

The number of hierarchical layers and the number of communities in each hierarchical layer are presented in Table \ref{tab:6}.

\begin{table}[htbp]
\caption{Structure of hierarchy and communities in each hierarchical layer for nine URNs.}
\label{tab:6}       % Give a unique label
% resize the table to not extend the text width
\resizebox{\textwidth}{!}{
\begin{tabular}{cccc}
\hline\noalign{\smallskip}
\textbf{City} & \textbf{Nodes} & \textbf{Hierarchy layers} &  \textbf{\tabincell{c}{Communities in each \\ hierarchy(from low to high)}} \\
\noalign{\smallskip}\hline\noalign{\smallskip}
\textbf{Austin} & 7388 & 5 & 151; 46; 13; 7; 5 \\
\textbf{Chicago} & 12982 & 5 & 266; 81; 22; 7; 2 \\
\textbf{Philadelphia} & 13389 & 6 & 343; 100; 29; 11; 6; 3 \\
\textbf{Anaheim} & 416 & 2 & 11; 2 \\
\textbf{Winnipeg} & 1052 & 3 & 30; 8; 7 \\
\textbf{Berlin} & 12981 & 5 & 457; 112; 34; 14; 5 \\
\textbf{Barcelona} & 1020 & 3 & 20; 6; 4 \\
\textbf{Terrassa} &1609 & 3 & 27; 5; 3 \\
\textbf{Hessen} & 4660 & 4 & 104; 25; 8; 5 \\
\noalign{\smallskip}\hline
\end{tabular}}
\end{table}

Identifying the modules and structure of the hierarchy using the topology information encoded in the networks is the main purpose of detecting communities (Fortunato \citeyear{fortunato2010community}). As can be seen from Table \ref{tab:6}, larger URNs are more likely to have more communities and hierarchical layers. For example, Philadelphia, Berlin and Chicago have 6, 5 and 5 hierarchy layers respectively, and the number of communities in the lowest layers is 343, 457 and 266 respectively. Anaheim, with just 416 nodes, however, has just two hierarchical layers and the number of communities at the lowest layer is just 11.

\begin{figure}[htbp]
\centering
\includegraphics[width=1\textwidth]{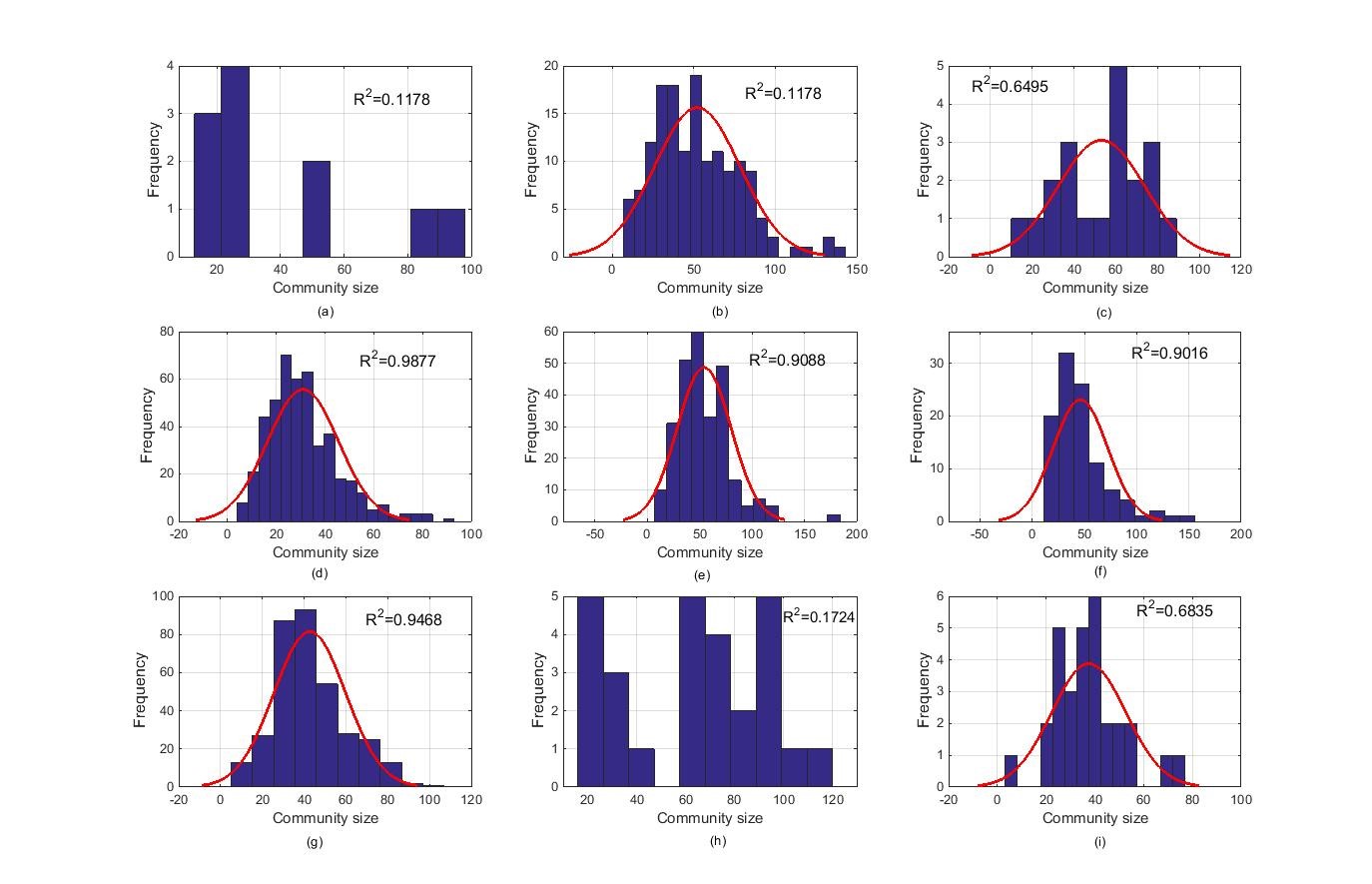}
\caption{Histograms for nine URNs: a. Anaheim; b. Austin; C. Barcelona; d. Berlin; e. Chicago; f. Hessen; g. Philadelphia; h. Terrassa; i. Winnipeg.}
\label{fig:11}       % Give a unique label
\end{figure}

All these URNs consist of many communities, with each community consisting of many individual nodes, and hence the size of communities varies. To some extent, these communities may follow a certain distribution rule. In order to identify this point, histograms can be plotted, as shown in Fig. \ref{fig:11}.

As can be seen from Fig. \ref{fig:11}, the larger URNs, such as Austin, Berlin and Chicago, tend to follow a normal distribution. Among these URNs, the left side of sub-figures regarding Austin, Berlin and Chicago does not fit well with normal distribution compared to their right side. The red lines in the graphs are the normal distribution fitting lines and clearly identify this point. $R^2$, meanwhile, which is calculated based on the maximum likelihood method, represents the goodness of the communities of the URNs following normal distribution. Conversely, those URNs with a smaller size apparently have worse normal distribution, and it seems that Anaheim and Winnipeg do not follow any distribution rules due to the fact that these URNs are relatively small so they are not too many hierarchical layers and communities. Through the above analysis, a rough conclusion is able to be summarised; that is, the larger the URNs are the better the normal distribution fitting they have.

The detection of communities for URNs may provide insightful views for the management and planning of transport within urban areas. For example, installing detectors or any other electronic facilities based on the distribution of communities may efficiently collect data such as traffic flows, driver behaviours and traffic incidents to improve network safety or mitigate traffic congestion. In addition, identification of ``sharing nodes'' is helpful for protecting critical parts of the systems and maintaining robustness of URNs. Furthermore, through the detection and analysis to communities and hierarchies of URNs, we may say that the URNs also tend to follow general nature rules in natural systems(Casella and Berger \citeyear{casella2002statistical}), such as a hierarchical structure and a normal distribution.

\section{Conclusions}
\label{conclusions}

As technologies related to smart city and cooperative vehicle infrastructure systems (CVIS) is increasingly prevailing, urban roads and its constituent urban road networks receive more and more attention. In this context, this paper has investigated the topological characteristics of a number of URNs for the first time in terms of purely physical roads. Widely used topology indices based on complex network theory were used in this paper to explore the topological characteristics of the nine URNs, small-world analysis and correlation analysis between some topological indices are conducted, and relatively new notions such as community detection were also applied to supplement the understanding of these topological networks. The results can be summarised as follows:

\begin{itemize}
         \item The average indegree and average outdegree of these URNs are exactly the same, and the distribution of indegree and outdegree are very similar, despite the fact that these URNs are directional. In addition, the distributions of indegree and outdegree for these URNs follow power law, showing that most of nodes have small indegree and outdegree, which means the URNs overall are less well connected. Philadelphia is overall less well connected than the other URNs.
         \item The clustering coefficients $(CC)$ of the URNs are very low, suggesting the URNs are less compact in local areas. The average CC of these URNs is smaller compared to other transport networks.
         \item Most of the weighted closeness of these URNs also follows an exponential distribution, and Anaheim and Berlin are the URNs in which most of the nodes have worse accessibility to other nodes.
         \item The weighted betweenness centrality (WBC) of all URNs follows a power law distribution well, and the parameter $\lambda (P(BC \geq x) \propto x^{-\lambda})$ for the distributions is between 0.4 and 0.8.
         \item $APL$ and diameters are proportional with the size of the URNs, and efficiency measures are consistent with APL. The results show that Berlin has the worst efficiency, while Anaheim has the best efficiency in terms of $APL$.
         \item Unlike other networked systems, such as metabolic networks and transport networks, these URNs do not have significant small-world properties.
\end{itemize}

These results reveal that URNs have distinctive statistical characteristics compared to other transport networks, such as the Chinese airport network (Li and Cai \citeyear{li2004statistical}), the Italian airport network (Guida and Maria \citeyear{guida2007topology}), the world-wide air transportation network (Guimera and Mossa et al. \citeyear{guimera2005worldwide}) and the Boston subway network (Latora and Marchiorib \citeyear{latora2002boston}), particularly these URNs do not have significant small-worldness, which may be caused by the unique spatial properties of urban roads.

Based on this, the relationships among indegree, outdegree, \textsl{WBC} and $CC$ have been examined, the results reveal that the correlations between degree, \textsl{WBC} and $CC$ are not significant; neither highly positive relationships nor anomalies exist in these URNs.

Following this, community detections for the URNs were conducted using the OSLOM method, and the results show that the larger the URNs were, the more communities and hierarchical layers they had. In addition to this, the distributions of communities at the lowest hierarchical layer for the URNs tend to fit a normal distribution better if the URNs are larger. These findings are compatible with common phenomena in natural and social science (Casella and Berger \citeyear{casella2002statistical}). The community detection of the URNs will facilitate the detector installation, data collection and identification of critical parts of cities.

The purpose of this study has been to explore the unique characteristics of the URNs. This work also helps to prepare the ground for further study of other important properties of URNs, such as robustness and resilience. The topological indices shown in this study are usually related to the robustness of networks due to the fact that some of these measures are able to reflect the efficiency and functionality of those networks (Albert and Jeong et al. \citeyear{albert2000error}, Holme and Kim et al. \citeyear{holme2002attack}, Crucitti and Latora et al. \citeyear{crucitti2003efficiency}, Crucitti and Latora et al. \citeyear{crucitti2004error}), and this feature can be used for the assessment of robustness and resilience combining the operational data of URNs, such as the traffic flow and road capacities.

Due to the fact that the data sources related to URNs are very limited, in this study, the URNs are mainly located in Europe and North America, such as Chicago, Austin, Berlin and Barcelona. In the future, with data sourced from other parts of the world, in particular from less developed countries or regions, it will be possible to expand our comparisons to a wider context. In particular, it is hypothesized that the diversity in social, economic, cultural and geographical characteristics could manifest itself in the formation of URNs, and thus may be reflected in the topological features of these URNs. Future work along these lines may uncover qualitatively different topological or operational features of URNs from a geographically representative set of test networks.

%\begin{acknowledgements}
%\end{acknowledgements}

\bibliographystyle{spbasic}      % spbasic basic style, author-year citations
\bibliography{complexnetwork}   % name your BibTeX data base

\begin{thebibliography}{52}
\providecommand{\natexlab}[1]{#1}
\providecommand{\url}[1]{{#1}}
\providecommand{\urlprefix}{URL }
\expandafter\ifx\csname urlstyle\endcsname\relax
  \providecommand{\doi}[1]{DOI~\discretionary{}{}{}#1}\else
  \providecommand{\doi}{DOI~\discretionary{}{}{}\begingroup
  \urlstyle{rm}\Url}\fi
\providecommand{\eprint}[2][]{\url{#2}}

\bibitem[{Albert et~al.(2000)Albert, Jeong, and Barab{\'a}si}]{albert2000error}
Albert R, Jeong H, Barab{\'a}si AL (2000) Error and attack tolerance of complex
  networks. nature 406(6794):378

\bibitem[{Barab{\'a}si and Albert(1999)}]{barabasi1999emergence}
Barab{\'a}si AL, Albert R (1999) Emergence of scaling in random networks.
  science 286(5439):509--512

\bibitem[{Bellingeri et~al.(2018)Bellingeri, Bevacqua, Scotognella, Zhe-Ming,
  and Cassi}]{bellingeri2018efficacy}
Bellingeri M, Bevacqua D, Scotognella F, Zhe-Ming L, Cassi D (2018) Efficacy of
  local attack strategies on the beijing road complex weighted network. Physica
  A: Statistical Mechanics and its Applications 510:316--328

\bibitem[{B{\'o}ta et~al.(2017)B{\'o}ta, Gardner, and
  Khani}]{bota2017identifying}
B{\'o}ta A, Gardner LM, Khani A (2017) Identifying critical components of a
  public transit system for outbreak control. Networks and Spatial Economics
  17(4):1137--1159

\bibitem[{Casella and Berger(2002)}]{casella2002statistical}
Casella G, Berger RL (2002) Statistical inference, vol~2. Duxbury Pacific
  Grove, CA

\bibitem[{Crucitti et~al.(2003)Crucitti, Latora, Marchiori, and
  Rapisarda}]{crucitti2003efficiency}
Crucitti P, Latora V, Marchiori M, Rapisarda A (2003) Efficiency of scale-free
  networks: error and attack tolerance. Physica A: Statistical Mechanics and
  its Applications 320:622--642

\bibitem[{Crucitti et~al.(2004)Crucitti, Latora, Marchiori, and
  Rapisarda}]{crucitti2004error}
Crucitti P, Latora V, Marchiori M, Rapisarda A (2004) Error and attack
  tolerance of complex networks. Physica A: Statistical mechanics and its
  applications 340(1-3):388--394

\bibitem[{Derrible(2012)}]{derrible2012network}
Derrible S (2012) Network centrality of metro systems. PloS one 7(7):e40575

\bibitem[{Dimitrov and Ceder(2016)}]{dimitrov2016method}
Dimitrov SD, Ceder AA (2016) A method of examining the structure and
  topological properties of public-transport networks. Physica A: Statistical
  Mechanics and its Applications 451:373--387

\bibitem[{Duan and Lu(2014)}]{duan2014robustness}
Duan Y, Lu F (2014) Robustness of city road networks at different
  granularities. Physica A: Statistical Mechanics and its Applications
  411:21--34

\bibitem[{Ducruet and Beauguitte(2014)}]{Ducruet2014complex}
Ducruet C, Beauguitte L (2014) Spatial science and network science: Review and
  outcomes of a complex relationship. Networks and Spatial Economics
  14(3):297--316, \doi{10.1007/s11067-013-9222-6},
  \urlprefix\url{https://doi.org/10.1007/s11067-013-9222-6}

\bibitem[{von Ferber et~al.(2007)von Ferber, Holovatch, Holovatch, and
  Palchykov}]{von2007network}
von Ferber C, Holovatch T, Holovatch Y, Palchykov V (2007) Network harness:
  Metropolis public transport. Physica A: Statistical Mechanics and its
  Applications 380:585--591

\bibitem[{von Ferber et~al.(2009)von Ferber, Holovatch, Holovatch, and
  Palchykov}]{von2009public}
von Ferber C, Holovatch T, Holovatch Y, Palchykov V (2009) Public transport
  networks: empirical analysis and modeling. The European Physical Journal B
  68(2):261--275

\bibitem[{Fortunato(2010)}]{fortunato2010community}
Fortunato S (2010) Community detection in graphs. Physics reports
  486(3-5):75--174

\bibitem[{Goh et~al.(2016)Goh, Choi, Lee, and Kim}]{goh2016complexity}
Goh S, Choi M, Lee K, Kim Km (2016) How complexity emerges in urban systems:
  Theory of urban morphology. Physical Review E 93(5):052309

\bibitem[{Guida and Maria(2007)}]{guida2007topology}
Guida M, Maria F (2007) Topology of the italian airport network: A scale-free
  small-world network with a fractal structure? Chaos, Solitons \& Fractals
  31(3):527--536

\bibitem[{Guimera and Amaral(2004)}]{guimera2004modeling}
Guimera R, Amaral LAN (2004) Modeling the world-wide airport network. The
  European Physical Journal B 38(2):381--385

\bibitem[{Guimera et~al.(2005)Guimera, Mossa, Turtschi, and
  Amaral}]{guimera2005worldwide}
Guimera R, Mossa S, Turtschi A, Amaral LN (2005) The worldwide air
  transportation network: Anomalous centrality, community structure, and
  cities' global roles. Proceedings of the National Academy of Sciences
  102(22):7794--7799

\bibitem[{Guimer{\`a} et~al.(2007)Guimer{\`a}, Sales-Pardo, and
  Amaral}]{guimera2007network}
Guimer{\`a} R, Sales-Pardo M, Amaral LAN (2007) A network-based method for
  target selection in metabolic networks. Bioinformatics 23(13):1616--1622

\bibitem[{Guo and Cai(2008)}]{guo2008degree}
Guo L, Cai X (2008) Degree and weighted properties of the directed china
  railway network. International Journal of Modern Physics C 19(12):1909--1918

\bibitem[{Holme et~al.(2002)Holme, Kim, Yoon, and Han}]{holme2002attack}
Holme P, Kim BJ, Yoon CN, Han SK (2002) Attack vulnerability of complex
  networks. Physical review E 65(5):056109

\bibitem[{Hu and Zhu(2009)}]{hu2009empirical}
Hu Y, Zhu D (2009) Empirical analysis of the worldwide maritime transportation
  network. Physica A: Statistical Mechanics and its Applications
  388(10):2061--2071

\bibitem[{Illenberger et~al.(2013)Illenberger, Nagel, and
  Fl{\"o}tter{\"o}d}]{Illenberger2013}
Illenberger J, Nagel K, Fl{\"o}tter{\"o}d G (2013) The role of spatial
  interaction in social networks. Networks and Spatial Economics
  13(3):255--282, \doi{10.1007/s11067-012-9180-4},
  \urlprefix\url{https://doi.org/10.1007/s11067-012-9180-4}

\bibitem[{Kaluza et~al.(2010)Kaluza, K{\"o}lzsch, Gastner, and
  Blasius}]{kaluza2010complex}
Kaluza P, K{\"o}lzsch A, Gastner MT, Blasius B (2010) The complex network of
  global cargo ship movements. Journal of the Royal Society Interface
  7(48):1093--1103

\bibitem[{Karinthy(1929)}]{karinthy1929chains}
Karinthy F (1929) Chains. everything is different

\bibitem[{Krishnamurthy and Wang(2000)}]{krishnamurthy2000network}
Krishnamurthy B, Wang J (2000) On network-aware clustering of web clients. ACM
  SIGCOMM Computer Communication Review 30(4):97--110

\bibitem[{Lancichinetti et~al.(2011)Lancichinetti, Radicchi, Ramasco, and
  Fortunato}]{lancichinetti2011finding}
Lancichinetti A, Radicchi F, Ramasco JJ, Fortunato S (2011) Finding
  statistically significant communities in networks. PloS one 6(4):e18961

\bibitem[{Latora and Marchiori(2002)}]{latora2002boston}
Latora V, Marchiori M (2002) Is the boston subway a small-world network?
  Physica A: Statistical Mechanics and its Applications 314(1-4):109--113

\bibitem[{Li and Cai(2004)}]{li2004statistical}
Li W, Cai X (2004) Statistical analysis of airport network of china. Physical
  Review E 69(4):046106

\bibitem[{Li and Cai(2007)}]{li2007empirical}
Li W, Cai X (2007) Empirical analysis of a scale-free railway network in china.
  Physica A: Statistical Mechanics and its Applications 382(2):693--703

\bibitem[{Lozano and Guti{\'e}rrez(2011)}]{Lozano2011airports}
Lozano S, Guti{\'e}rrez E (2011) Efficiency analysis and target setting of
  spanish airports. Networks and Spatial Economics 11(1):139--157,
  \doi{10.1007/s11067-008-9096-1},
  \urlprefix\url{https://doi.org/10.1007/s11067-008-9096-1}

\bibitem[{Marshall et~al.(2018)Marshall, Gil, Kropf, Tomko, and
  Figueiredo}]{marshall2018street}
Marshall S, Gil J, Kropf K, Tomko M, Figueiredo L (2018) Street network
  studies: from networks to models and their representations. Networks and
  Spatial Economics \doi{10.1007/s11067-018-9427-9},
  \urlprefix\url{https://doi.org/10.1007/s11067-018-9427-9}

\bibitem[{Masucci et~al.(2014)Masucci, Stanilov, and
  Batty}]{masucci2014exploring}
Masucci AP, Stanilov K, Batty M (2014) Exploring the evolution of london's
  street network in the information space: A dual approach. Physical Review E
  89(1):012805

\bibitem[{Neal(2018)}]{neal2018urban}
Neal Z (2018) Is the urban world small? the evidence for small world structure
  in urban networks. Networks and Spatial Economics
  \doi{10.1007/s11067-018-9417-y},
  \urlprefix\url{https://doi.org/10.1007/s11067-018-9417-y}

\bibitem[{Newman(2003)}]{newman2003structure}
Newman ME (2003) The structure and function of complex networks. SIAM review
  45(2):167--256

\bibitem[{Newman and Peixoto(2015)}]{newman2015generalized}
Newman ME, Peixoto TP (2015) Generalized communities in networks. Physical
  review letters 115(8):088701

\bibitem[{Petri et~al.(2013)Petri, Expert, Jensen, and
  Polak}]{petri2013entangled}
Petri G, Expert P, Jensen HJ, Polak JW (2013) Entangled communities and spatial
  synchronization lead to criticality in urban traffic. Scientific reports
  3:1798

\bibitem[{Reddy et~al.(2002)Reddy, Kitsuregawa, Sreekanth, and
  Rao}]{reddy2002graph}
Reddy PK, Kitsuregawa M, Sreekanth P, Rao SS (2002) A graph based approach to
  extract a neighborhood customer community for collaborative filtering. In:
  International Workshop on Databases in Networked Information Systems,
  Springer, pp 188--200

\bibitem[{Sabidussi(1966)}]{sabidussi1966centrality}
Sabidussi G (1966) The centrality index of a graph. Psychometrika
  31(4):581--603

\bibitem[{Sienkiewicz and Ho{\l}yst(2005)}]{sienkiewicz2005statistical}
Sienkiewicz J, Ho{\l}yst JA (2005) Statistical analysis of 22 public transport
  networks in poland. Physical Review E 72(4):046127

\bibitem[{de~Sola~Pool and Kochen(1978)}]{de1978contacts}
de~Sola~Pool I, Kochen M (1978) Contacts and influence. Social networks
  1(1):5--51

\bibitem[{Sun et~al.(2018)Sun, Huang, Chen, and Yao}]{sun2018vulnerability}
Sun L, Huang Y, Chen Y, Yao L (2018) Vulnerability assessment of urban rail
  transit based on multi-static weighted method in beijing, china.
  Transportation Research Part A: Policy and Practice 108:12--24

\bibitem[{Sun et~al.(2014)Sun, Wandelt, and Linke}]{sun2014topological}
Sun X, Wandelt S, Linke F (2014) Topological properties of the air navigation
  route system using complex network theory. In: 6th international conference
  on research in air transportation (ICRAT2014)

\bibitem[{Travers and Milgram(1967)}]{travers1967small}
Travers J, Milgram S (1967) The small world problem. Phychology Today
  1(1):61--67

\bibitem[{Tsiotas and Polyzos(2015)}]{tsiotas2015analyzing}
Tsiotas D, Polyzos S (2015) Analyzing the maritime transportation system in
  greece: a complex network approach. Networks and Spatial Economics
  15(4):981--1010

\bibitem[{V{\'a}zquez et~al.(2002)V{\'a}zquez, Pastor-Satorras, and
  Vespignani}]{vazquez2002large}
V{\'a}zquez A, Pastor-Satorras R, Vespignani A (2002) Large-scale topological
  and dynamical properties of the internet. Physical Review E 65(6):066130

\bibitem[{Watts and Strogatz(1998)}]{watts1998collective}
Watts DJ, Strogatz SH (1998) Collective dynamics of ‘small-world’networks.
  nature 393(6684):440

\bibitem[{Wu and Huberman(2004)}]{wu2004finding}
Wu F, Huberman BA (2004) Finding communities in linear time: a physics
  approach. The European Physical Journal B 38(2):331--338

\bibitem[{Xu et~al.(2007)Xu, Hu, Liu, and Liu}]{xu2007scaling}
Xu X, Hu J, Liu F, Liu L (2007) Scaling and correlations in three bus-transport
  networks of china. Physica A: Statistical Mechanics and its Applications
  374(1):441--448

\bibitem[{Zhan et~al.(2017)Zhan, Ukkusuri, and Rao}]{zhan2017dynamics}
Zhan X, Ukkusuri SV, Rao PSC (2017) Dynamics of functional failures and
  recovery in complex road networks. Physical Review E 96(5):052301

\bibitem[{Zhang et~al.(2016)Zhang, Derudder, and Witlox}]{zhang2016dynamics}
Zhang S, Derudder B, Witlox F (2016) Dynamics in the european air transport
  network, 2003--9: an explanatory framework drawing on stochastic actor-based
  modeling. Networks and Spatial Economics 16(2):643--663

\bibitem[{Zheng et~al.(2012)Zheng, Chen, Shao, and Bie}]{zheng2012analysis}
Zheng X, Chen JP, Shao JL, Bie LD (2012) Analysis on topological properties of
  beijing urban public transit based on complex network theory. Acta Physica
  Sinica 61:190510

\end{thebibliography}

%\begin{thebibliography}{}
% etc
%\end{thebibliography}

\end{document}